\documentclass[%
 reprint,
 amsmath,amssymb,
 aps,
 prd,
]{revtex4-2}

\usepackage{graphicx}% Include figure files
\usepackage{dcolumn}% Align table columns on decimal point
\usepackage{bm}% bold math
\usepackage{amsfonts}
\usepackage{bbm}
\usepackage{comment}
\usepackage{hyperref}% add hypertext capabilities
\usepackage{orcidlink}
\def\pd{{\rm d}}
\def\mpc{{\rm Mpc}}

\def\pa#1#2{\dfrac{\partial #1}{\partial #2}}

\def\deltadir{\delta_{\rm D}}

\def\expt{\mathbbm{E}}

\newcommand\CG[1]{\textcolor{black}{#1}}

\begin{document}

\preprint{APS/123-QED}

\title{Wasserstein Distance in Cosmological Structure Formation: An Optimal Transport Perspective}% Force line breaks with \\

\author{Tsutomu T.\ Takeuchi\,\orcidlink{0000-0001-8416-7673}}
 \email{tsutomu.takeuchi.ttt@gmail.com}
\affiliation{%
 Division of Particle and Astrophysical Science, Nagoya University, \\
 Furo-cho, Chikusa-ku, Nagoya, 464–8602, Japan\\
}%
\affiliation{
 The Research Center for Statistical Machine Learning, the Institute of Statistical Mathematics, \\
 10-3 Midori-cho, Tachikawa, Tokyo 190–8562, Japan
}%

\date{\today}% It is always \today, today,

\begin{abstract}
The formation of cosmological large-scale structure is usually described in terms of the evolution of density fluctuations and their statistical measures, such as the power spectrum and correlation function.
However, these statistics characterize the amplitude structure of density fluctuations and do not directly describe the spatial redistribution of matter that occurs during structure formation.
In this work we formulate cosmological structure formation as a transport problem of mass distributions using the Wasserstein distance from optimal transport theory.
The generative process from the initial linear density field to the observed galaxy catalog is treated as a hierarchical mapping from a continuous density field to a galaxy point process, and an approximate expression for the Wasserstein distance between them is derived under the small-fluctuation approximation.
We show that this distance naturally decomposes into contributions associated with three physical processes: mass transport by gravitational evolution, galaxy formation bias, and shot noise arising from the discrete sampling of galaxies.
The gravitational transport term is expressed as an integral of the matter power spectrum, while the galaxy formation contribution appears as a transport-weighted integral of the galaxy correlation function.
The sampling term corresponds to Poisson shot noise originating from the discreteness of the galaxy catalog.
A notable feature of the resulting expression is that the correlation contribution differs fundamentally from conventional volume-weighted clustering statistics and is naturally related to cumulative matter displacements and baryon acoustic oscillation (BAO) smearing.
These results provide a unified framework for describing cosmological large-scale structure formation from the perspective of transport geometry and suggest that the Wasserstein distance may serve as a new statistical quantity linking continuous density fields with observed galaxy catalogs.
\end{abstract}

\maketitle

%\tableofcontents

\section{Introduction}

The formation of cosmological large-scale structure is understood as the growth of primordial density fluctuations driven by gravitational instability.
This process is one of the central problems of cosmology, and the statistical properties of the galaxy distribution and the cosmic web have traditionally been described using the power spectrum and correlation function \citep{1980lssu.book.....P,takeuchi2025physics}.
The growth of the initial fluctuations and structure formation in the weakly nonlinear regime are known to be effectively described by Lagrangian perturbation theory, represented by the Zel'dovich approximation \citep{1970A&A.....5...84Z}.
In this framework the mapping from Lagrangian coordinates to Eulerian coordinates is given by the particle displacement field, and cosmological structure formation can therefore be viewed not only as the growth of density fluctuations but also as a large-scale redistribution of matter through particle displacements.

\CG{From this perspective, optimal transport theory provides a natural mathematical framework for describing structure formation as a transport process.
The quadratic Wasserstein distance quantifies the minimal cost required to rearrange one mass distribution into another \citep{villani2009optimal}.
Optimal-transport methods have recently attracted considerable attention in cosmology, particularly in reconstruction problems that seek to recover the initial conditions from the observed galaxy distribution \citep{2002Natur.417..260F,2022PhRvL.129y1101N,2023PhRvD.108h3534N,2024PhRvD.109l3512N}.}

\CG{Most existing cosmological applications of optimal transport have focused on reconstruction problems, in which the goal is to infer the initial mass distribution from the observed large-scale structure.}
\CG{In contrast, the objective of the present work is to use transport geometry itself as a statistical descriptor of cosmological structure formation.}
\CG{The cosmological mass distribution evolves from the initial linear density field to a nonlinear density field through gravitational evolution and is subsequently observed as a finite set of galaxy positions after galaxy formation.}
\CG{We therefore consider the entire formation process as a sequence of mappings connecting the primordial density field to the observed galaxy catalog.}
The observed galaxy catalog can therefore be conceptually regarded as arising from a hierarchical generative process
\begin{align}
\rho_{\rm lin}
\longrightarrow
\rho_{\rm NL}
\longrightarrow
\lambda_{\rm gal}
\longrightarrow
\mu_N .
\end{align}
Here $\rho_{\rm lin}$ denotes the initial linear density field, $\rho_{\rm NL}$ the nonlinear density field after gravitational evolution, $\lambda_{\rm gal}$ the galaxy formation intensity field, and $\mu_N$ the point-process measure representing the observed galaxy catalog.

The goal of this work is to describe this entire generative process from the perspective of optimal transport distances.
In particular, we consider the Wasserstein distance between the initial density field and the observed galaxy catalog and analyze its statistical properties under the small-fluctuation approximation.
\CG{Our central result is that} this distance naturally decomposes into contributions associated with three physical processes.
Specifically,
\begin{align}
W_2^2
\simeq
W_{\rm grav}^2
+
W_{\rm bias}^2
+
W_{\rm samp}^2 ,
\end{align}
which correspond respectively to gravitational evolution, galaxy formation bias, and shot-noise effects arising from the finite sampling of galaxies.

As the main result of this study, the statistical mean of the Wasserstein distance between the initial density field and the observed galaxy catalog can be approximately written as
\begin{align}
\left\langle W_2^2 \right\rangle
\simeq
\frac{1}{2\pi^2}
\int_0^\infty
P_m(k),\pd k
+
\int_0^\infty
r,\xi_{\rm gal}(r),\pd r
+
\frac{1}{4\pi^{3/2}\bar n R} .
\end{align}
\CG{The first term represents gravitational transport, the second describes the correlation contribution of the observed galaxy distribution, and the third corresponds to shot noise arising from finite sampling.}
\CG{A particularly noteworthy aspect of this result is that the correlation contribution appears in the transport-weighted form}
\begin{align}
\CG{\Delta W_2^2\sim\int r,\xi(r),\pd r .}
\end{align}
\CG{Unlike conventional volume-weighted clustering measures, this quantity depends linearly on separation and is therefore sensitive to cumulative mass displacements rather than pair counts alone.}
\CG{Because large-scale coherent displacements are also responsible for the smearing of the baryon acoustic oscillation (BAO) feature, this transport interpretation suggests a direct connection between Wasserstein geometry and observable signatures of large-scale structure formation.}

The structure of this paper is as follows.
\CG{
Section~\ref{sec:point_process_galaxies} introduces the point-process description of galaxy distributions and the associated statistical framework. Sections~\ref{sec:wasserstein_cosmology} and \ref{sec:point_process_limit_wasserstein} develop the optimal-transport framework and establish the correspondence between continuous density fields and galaxy catalogs.
Sections~\ref{sec:density_fluctuation_wasserstein} and \ref{sec:finite_volume_effect} derive the Wasserstein distance for cosmological density fluctuations and investigate the effects of finite observational volume.
Sections~\ref{sec:galaxy_formation_as_point_process}, \ref{sec:correlated_process_wasserstein}, and \ref{sec:unified_galaxy_bias_wasserstein} extend the formalism to galaxy formation, stochastic sampling, correlated point processes, and galaxy bias, leading to a unified description of the Wasserstein distance for observed galaxy catalogs.
Finally, Sec.~\ref{sec:discusssion_conclusion} discusses the physical implications of the results and presents conclusions and future prospects.
}

\section{Point-process description of galaxy distributions}
\label{sec:point_process_galaxies}

\CG{Cosmological observations considered in this work consist of the positions of a finite number of galaxies or dark matter halos,}
\begin{align}
X=\{\bm{x}_i\}_{i=1}^{N} .
\end{align}
\CG{where $\bm{x}_i\in W$ and the observational region satisfies}
\begin{align}
W\subset\mathbb{R}^3 .
\end{align}
\CG{Such data are naturally described as realizations of a spatial point process}
\citep{1980lssu.book.....P,takeuchi2025physics}.
\CG{The purpose of this section is to formulate galaxy catalogs as measures so that they can be treated within the framework of optimal transport theory.}

\subsection{Galaxy distributions as discrete measures}

The finite point set $X$ can be represented measure-theoretically as a discrete measure.
Assigning a mass $m_i$ to each galaxy, the galaxy distribution is written as
\begin{align}
\mu
=
\sum_{i=1}^{N} m_i\,\deltadir(\bm{x}-\bm{x}_i) .
\end{align}
Here $\deltadir$ denotes the Dirac delta function.
\CG{This representation treats the galaxy catalog as a discrete sampling of the underlying cosmological mass distribution.}

\subsection{Relation to the continuous density field}

\CG{In theoretical cosmology, the mass distribution is represented by a continuous density field $\rho(\bm{x})$, corresponding to the measure}
\begin{align}
\mu(\pd^3 x )
=
\rho(\bm{x})\,\pd^3 x .
\end{align}
\CG{The observed galaxy catalog may therefore be viewed as a discrete approximation to an underlying continuous measure,}
\begin{align}
\rho(\bm{x})\,\pd^3 x
\;\longrightarrow\;
\sum_{i=1}^{N}\deltadir(\bm{x}-\bm{x}_i) .
\end{align}
\CG{The aim of this work is to describe this mapping within the framework of optimal transport theory.}
In optimal transport theory, measures are usually defined as probability measures.
Accordingly, we normalize the density field by the total mass
\begin{align}
M
=
\int \rho(\bm{x})\,\pd^3 x .
\end{align}
and consider the probability density
\begin{align}
\tilde{\rho}(\bm{x})
=
\frac{\rho(\bm{x})}{M} .
\end{align}
The associated probability measure is
\begin{align}
\tilde{\mu}(\pd^3 x )
=
\tilde{\rho}(\bm{x})\,\pd^3 x .
\end{align}
\CG{This defines an element of $\mathcal P_2(\mathbb{R}^3)$, the space of probability measures on $\mathbb{R}^3$ with finite second moments, which is the natural domain of the quadratic Wasserstein distance.}
In what follows, for notational simplicity, we again denote the normalized measure by $\mu$.

\section{Wasserstein distance and cosmological density fields}
\label{sec:wasserstein_cosmology}

\CG{Optimal transport theory equips the space of probability measures with a natural metric, the Wasserstein distance, which quantifies the cost of rearranging one mass distribution into another \citep[e.g.,][]{villani2009optimal}.}
\CG{Because cosmology involves both continuous density fields and discrete galaxy catalogs, the Wasserstein distance provides a natural framework for treating these objects within a unified geometric setting.}

\subsection{Quadratic Wasserstein distance}

For two probability measures
\begin{align}
\mu,\nu\in\mathcal{P}_2(\mathbb{R}^3) ,
\end{align}
the quadratic Wasserstein distance associated with the squared-distance cost
\begin{align}
c(\bm{x},\bm{y}) = |\bm{x}-\bm{y}|^2
\end{align}
is defined by
\begin{align}
W_2^2(\mu,\nu)
=
\inf_{\gamma\in\Pi(\mu,\nu)}
\int |\bm{x}-\bm{y}|^2\,\pd\gamma(\bm{x},\bm{y}) .
\end{align}
\CG{Here $\Pi(\mu,\nu)$ denotes the set of transport plans, i.e., joint probability measures on $\mathbb{R}^3\times\mathbb{R}^3$ whose marginals are $\mu$ and $\nu$.}
\CG{A transport plan specifies how mass elements in $\mu$ are redistributed to form $\nu$.}
When an optimal map $T$ exists, this can be written as
\begin{align}
W_2^2(\mu,\nu)
=
\int |\bm{x}-T(\bm{x})|^2\,\pd\mu(\bm{x}) .
\end{align}

\subsection{Dynamic formulation}

\citet{BenamouBrenier2000} showed that the Wasserstein distance can be written as a hydrodynamic minimum-action principle.
\CG{Let $\mu_0$ and $\mu_1$ denote the initial and final probability measures, respectively.}
Introducing the density field $\rho(\bm{x},t)$ and the velocity field $\bm{v}(\bm{x},t)$, one has
\begin{align}
W_2^2(\mu_0,\mu_1)
=
\min_{\rho,\bm{v}}
\int_0^1
\int
\rho|\bm{v}|^2
\,\pd^3 x\,\pd t .
\end{align}
Here $\rho$ and $\bm{v}$ are required to satisfy the continuity equation
\begin{align}
\pa{\rho}{t}
+
\bm{\nabla}\cdot(\rho\bm{v})
=
0 .
\end{align}
\CG{This formulation is particularly relevant for cosmology because cold dark matter behaves approximately as a pressureless fluid on large scales.}

In the early stage of cosmological structure formation, cold dark matter can be approximated as a pressureless fluid, and its time evolution is described by the Euler--Poisson system
\begin{align}
\pa{\rho}{t}
+
\bm{\nabla}\cdot(\rho\bm{v})
&=0,\\
\pa{\bm{v}}{t}
+
(\bm{v}\cdot\bm{\nabla})\bm{v}
&=
-\bm{\nabla}\Phi,\\
\nabla^2\Phi
&=
4\pi G\rho .
\end{align}
Therefore, on large scales the velocity field is approximately a potential flow.
In this regime, the time evolution of the matter distribution can be approximately understood as geodesic motion in Wasserstein space.

\subsection{Zel'dovich approximation and Wasserstein geodesics}

In the early stage of cosmological structure formation, the motion of matter is described by the mapping from the Lagrangian coordinate $\bm{q}$ to the Eulerian coordinate $\bm{x}$,
\begin{align}
\bm{x}(\bm{q},t)
=
\bm{q}
+
\bm{\Psi}(\bm{q},t) .
\end{align}
Here $\bm{\Psi}$ is the displacement field.

In the linear regime, the Zel'dovich approximation \citep{1970A&A.....5...84Z} gives
\begin{align}
\bm{\Psi}(\bm{q},t)
=
-D(t)\bm{\nabla}\phi(\bm{q}) ,
\end{align}
where $D(t)$ is the linear growth factor and $\phi(\bm{q})$ is the initial gravitational potential.
Accordingly, the Eulerian coordinate is
\begin{align}
\bm{x}(\bm{q},t)
=
\bm{q}
-
D(t)\bm{\nabla}\phi(\bm{q}) .
\end{align}

\CG{Because the displacement field is generated by a scalar potential, the Zel'dovich mapping can be written as a gradient map}
\begin{align}
\bm{x}
=
\bm{\nabla}\Phi(\bm{q}) .
\end{align}

\CG{In the quadratic-cost optimal transport problem, Brenier's theorem states that the optimal transport map is also given by the gradient of a convex potential \citep{Brenier1991}.}
\CG{The Zel'dovich approximation can therefore be interpreted as a Wasserstein geodesic transporting the initial mass distribution to the evolved density field.}

\CG{Mass conservation implies that the evolved density field is determined by the Jacobian of the transformation between Lagrangian and Eulerian coordinates.}
\CG{Because this transformation is generated by a scalar potential, it possesses the same gradient-map structure that characterizes quadratic optimal transport.}

\CG{This correspondence provides the geometric basis for interpreting cosmological structure formation as a transport process in the space of probability measures and underlies several optimal-transport reconstruction methods in cosmology \citep{2002Natur.417..260F}.}

\section{Point-process limit in Wasserstein space} \label{sec:point_process_limit_wasserstein}

\subsection{Empirical measures and finite sampling}

For the measure corresponding to a continuous density field,
\begin{align}
\mu(\pd^3 x )
=
\rho(\bm{x})\,\pd^3 x ,
\end{align}
suppose that independent samples
\begin{align}
X
=
\{\bm{x}_i\}_{i=1}^{N}
\end{align}
are given.
One can then define the empirical measure
\begin{align}
\mu_N
=
\frac{1}{N}\sum_{i=1}^{N}\deltadir(\bm{x}-\bm{x}_i) .
\end{align}

\CG{The empirical measure provides a stochastic approximation to the underlying distribution $\mu$.}
Under suitable conditions, the empirical measure converges to $\mu$ with respect to the Wasserstein distance,
\begin{align}
W_2(\mu_N,\mu)\rightarrow 0
\quad (N\rightarrow\infty) .
\end{align}

The convergence rate depends on the spatial dimension $d$ and scales as
\begin{align}
\expt\bigl[W_2(\mu_N,\mu)\bigr]
\sim
N^{-1/d}
\end{align}
\citep{BoissardLeGouic2014,FournierGuillin2015}.

For cosmological applications, where $d=3$, one obtains
\begin{align}
\expt\bigl[W_2(\mu_N,\mu)\bigr]
\sim
N^{-1/3}.
\end{align}
\CG{This scaling implies that a finite galaxy catalog retains an intrinsic geometric sampling error with respect to the underlying continuous density field.}
\CG{Consequently, finite-sampling effects naturally contribute to the Wasserstein distance and must be taken into account when interpreting transport-based statistics.}

\subsection{Point-process sampling}

\CG{Galaxy catalogs are commonly modeled as realizations of spatial point processes.}
The simplest case is a Poisson point process.
For an intensity function $\lambda(\bm{x})$, the expected number of points in an arbitrary region $A$ satisfies
\begin{align}
\expt[N(A)]
=
\int_A \lambda(\bm{x})\,\pd^3 x
\end{align}
\citep{2003Daley_point_processI,2008Daley_point_processII,Last_Penrose_2017}.

\CG{Real galaxy distributions exhibit nontrivial clustering generated by gravitational evolution and galaxy formation.}
\CG{Such clustering is conventionally described by the correlation functions of the point process, most notably the two-point correlation function $\xi(r)$.}

\CG{From the present viewpoint, the observed galaxy catalog may be regarded as a stochastic discrete approximation to an underlying continuous density field,}
\begin{align}
\CG{
\rho(\bm{x})\,\pd^3x
\longrightarrow
\frac{1}{N}
\sum_{i=1}^{N}
\deltadir(\bm{x}-\bm{x}_i)
}.
\end{align}
\CG{The Wasserstein distance therefore contains contributions arising both from the continuous transport of matter and from the finite stochastic sampling that converts a continuous density field into a discrete galaxy catalog.}
\CG{The latter contribution will appear as a shot-noise term in the following analysis.}

\section{Wasserstein distance and cosmological density fluctuations}
\label{sec:density_fluctuation_wasserstein}

In cosmology the mass distribution is described by a continuous density field $\rho(\bm{x})$.
In the linear regime, where density fluctuations are small, the density field can be written around the mean density $\bar\rho$ as
\begin{align}
\rho(\bm{x})
=
\bar\rho\bigl(1+\delta(\bm{x})\bigr) .
\end{align}
Here $\delta(\bm{x})$ denotes the density fluctuation field.

Under the condition of mass conservation, the optimal transport map can be expressed in terms of a potential $\psi$ as
\begin{align}
T(\bm{x})
=
\bm{x}
+
\bm{\nabla}\psi(\bm{x}) .
\end{align}
When the density fluctuations are small, the transport equation can be linearized, yielding
\begin{align}
\nabla^2\psi
=
\delta .
\end{align}
This is a Poisson-type equation derived from the mass conservation condition for the normalized density field.
Under this approximation, the quadratic Wasserstein distance can be written as
\begin{align}
W_2^2
\simeq
\int
|\bm{\nabla}\psi|^2
\,\pd^3 x .
\end{align}
\CG{This expression establishes a direct connection between Wasserstein geometry and the displacement field responsible for cosmological structure formation.}

\subsection{Power-spectrum representation}

Expanding the density fluctuation field in Fourier space,
\begin{align}
\delta(\bm{x})
=
\int
\delta(\bm{k})
e^{i\bm{k}\cdot\bm{x}}
\frac{\pd^3 k}{(2\pi)^3} .
\end{align}
The potential is then written as
\begin{align}
\psi(\bm{k})
=
-\frac{\delta(\bm{k})}{k^2} .
\end{align}
Substituting this relation, the Wasserstein distance becomes
\begin{align}
W_2^2
=
\int
\frac{|\delta(\bm{k})|^2}{k^2}
\frac{\pd^3 k}{(2\pi)^3} .
\end{align}
The statistical properties of the density fluctuations are characterized by the power spectrum
\begin{align}
\langle
\delta(\bm{k})\delta^*(\bm{k}')
\rangle
=
(2\pi)^3
\deltadir(\bm{k}-\bm{k}')
P(k) .
\end{align}
Using this relation, the expectation value of the Wasserstein distance can be written as
\begin{align}
\langle W_2^2\rangle
=
\frac{1}{2\pi^2}
\int_0^\infty
P(k)\,\pd k .
\end{align}
\CG{Unlike conventional variance measures, which typically involve integrals weighted by powers of $k$, the Wasserstein distance depends directly on the integrated matter power spectrum.}
\CG{This reflects its geometric nature as a measure of cumulative mass displacement rather than fluctuation amplitude alone.}

\CG{\subsection{Physical interpretation: the Zel'dovich displacement field}}

\CG{Using the Zel'dovich displacement field introduced in Sec.~\ref{sec:wasserstein_cosmology},}
\begin{align}
\bm{\Psi}(\bm{k})
=
i\frac{\bm{k}}{k^2}\delta(\bm{k})
\end{align}
\citep{1970A&A.....5...84Z}.
Therefore the variance of the displacement field satisfies
\begin{align}
\langle |\bm{\Psi}|^2\rangle
\propto
\int
\frac{P(k)}{k^2}
\,\pd^3 k .
\end{align}

Using this relation, the Wasserstein distance associated with the evolution from the linear density field to the nonlinear density field can be written as
\begin{align}
W_2^2
=
\int
|\bm{\Psi}(\bm{q})|^2
\,\pd^3 q .
\end{align}

\CG{Thus the quadratic Wasserstein distance is directly proportional to the mean squared particle displacement.}
\CG{In this sense, the Wasserstein distance measures the cumulative transport of matter rather than the amplitude of density fluctuations themselves.}

\CG{Because large-scale coherent displacements are also responsible for bulk flows and for the damping of the baryon acoustic oscillation (BAO) feature, this interpretation provides a direct physical connection between Wasserstein geometry and observable signatures of structure formation.}
\CG{The implications of this connection will be discussed further in Sec.~\ref{sec:finite_volume_effect} and Sec.~\ref{sec:discusssion_conclusion}.}
\\

\section{Wasserstein geometry in a finite volume}
\label{sec:finite_volume_effect}

In the previous section we showed that, under the Zel'dovich approximation, the Wasserstein distance between the linear and nonlinear density fields can be expressed as the mean squared displacement field.
In particular, in the infinite-volume limit one has
\begin{align}
\langle W_2^2\rangle
=
\frac{1}{2\pi^2}
\int_0^\infty
P(k)\,\pd k .
\end{align}
In actual cosmological observations and numerical simulations, however, measurements are always performed in a finite volume
\begin{align}
V=L^3 .
\end{align}
It is well known in cosmology that a finite observational volume affects statistical quantities \citep{Peacock1999Cosmology,1994ApJ...426...23F}.
Here we discuss this finite-volume effect from the viewpoint of Wasserstein geometry.
\CG{Because long-wavelength displacements are also responsible for the damping of the baryon acoustic oscillation (BAO) feature, finite-volume effects have direct implications for observable signatures of large-scale structure formation.}

\subsection{Wasserstein distance in a finite volume}

Consider a cubic region with periodic boundary conditions,
\begin{align}
W=[0,L]^3 .
\end{align}
The allowed wave vectors are then
\begin{align}
\bm{k}
=
\frac{2\pi}{L}\bm{n},
\qquad
\bm{n}\in\mathbb{Z}^3 .
\end{align}
Accordingly, the minimum wave number is
\begin{align}
k_{\min}
=
\frac{2\pi}{L} .
\end{align}
From the result of the previous section, the Wasserstein distance in a finite volume can be written as
\begin{align}
W_{2,V}^2
=
\frac{1}{V}
\sum_{\bm{k}\neq\bm{0}}
\frac{|\delta(\bm{k})|^2}{k^2} .
\end{align}
Taking the statistical average gives
\begin{align}
\langle W_{2,V}^2\rangle
=
\frac{1}{V}
\sum_{\bm{k}\neq\bm{0}}
\frac{P(k)}{k^2} .
\end{align}

When the volume is sufficiently large, this sum can be approximated by
\begin{align}
\langle W_{2,V}^2\rangle
\simeq
\frac{1}{2\pi^2}
\int_{k_{\min}}^{k_{\max}}
P(k)\,\pd k .
\end{align}
Here $k_{\max}$ is the maximum wave number corresponding to the smoothing scale or numerical resolution.
\CG{This result shows that the Wasserstein distance is sensitive to long-wavelength modes and therefore depends explicitly on the observational volume through $k_{\min}$.}

\subsection{Long-wavelength modes and bulk flow}

\CG{The finite-volume dependence can be understood directly from the displacement-field interpretation of the Wasserstein distance.}
In a finite volume one has
\begin{align}
W_{2,V}^2
=
\frac{1}{V}
\int_W
|\bm{\Psi}(\bm{q})|^2
\,\pd^3 q .
\end{align}

As is clear from this expression, the finite-volume effect corresponds to the fact that the long-wavelength modes of the displacement field are not fully included in the observation.
In cosmology, such large-scale velocity fields are referred to as bulk flows \citep[e.g.,][]{1995PhR...261..271S}.
Within a finite observational region, long-wavelength modes that move the entire region almost uniformly are not adequately sampled.
\CG{This interpretation is closely related to the well-known damping of the BAO feature.}
\CG{Large-scale coherent displacements broaden the acoustic peak by moving galaxies away from their initial positions, thereby reducing the contrast of the BAO signal.
This effect has been extensively studied in the context of BAO damping and reconstruction \citep{2007ApJ...664..660E}.
The characteristic damping scale is determined by the variance of the displacement field, which is the same quantity that appears in the Wasserstein distance.}

\CG{From this perspective, the Wasserstein distance may be regarded as a transport-based measure of the cumulative displacement responsible for BAO smearing.}
\CG{This connection suggests a possible route toward relating Wasserstein geometry to observational constraints on the BAO damping scale.
Recent analyses have demonstrated that the BAO damping scale itself can be constrained observationally, providing a potentially direct link between Wasserstein-based statistics and measurable large-scale structure observables \citep{2026arXiv260214533P}.}

\subsection{Finite-window effects and observational geometry}

In actual galaxy surveys, data are obtained not with periodic boundary conditions but within a finite observational window of irregular shape,
\begin{align}
W\subset\mathbb{R}^3 .
\end{align}
The observed density field can be written using the window function $\mathcal{W}(\bm{x})$ as
\begin{align}
\delta_W(\bm{x})
=
\mathcal{W}(\bm{x})\delta(\bm{x}) .
\end{align}
In Fourier space this becomes
\begin{align}
\delta_W(\bm{k})
=
\int
\tilde{\mathcal{W}}(\bm{k}-\bm{k}')
\delta(\bm{k}')
\pd^3 k' ,
\end{align}
which induces mode mixing \citep{1994ApJ...426...23F}.
Therefore, the Wasserstein distance in a finite window is not simply given by an integral truncated at $k_{\min}$, but contains geometric effects determined by the shape of the observational window.
\CG{Consequently, the Wasserstein distance in a realistic survey depends not only on the finite volume but also on the geometry of the observational window through mode mixing.}

\section{Galaxy formation and point-process sampling}
\label{sec:galaxy_formation_as_point_process}

Up to this point we have formulated the evolution from the initial density field to the nonlinear matter distribution as a problem of transport geometry.
\CG{The remaining step is to connect the evolved density field to the observed galaxy catalog through galaxy formation and stochastic sampling.}

\CG{Let $\lambda_{\rm gal}(\bm{x})$ denote the galaxy formation intensity field and $\mu_N$ the empirical measure associated with the observed galaxy catalog.}
\CG{The mapping from the matter distribution to the observed galaxy sample may then be written schematically as}
\begin{align}
\mu_N
=
(\mathcal S\circ\mathcal B)(\rho_{\rm NL}) .
\end{align}
Here $\mathcal B$ denotes the bias operator that maps the density field to the galaxy formation intensity, and $\mathcal S$ denotes the sampling operator that maps the intensity field to a point process.
\CG{In this section we briefly discuss these two ingredients before incorporating them into the Wasserstein framework.}

\CG{\subsection{From density fields to galaxy catalogs}}

In the simplest model, the galaxy formation intensity can be written as a local function of the nonlinear density fluctuation $\delta(\bm{x})$,
\begin{align}
\lambda_{\rm gal}(\bm{x})
=
\bar n\,F(\delta(\bm{x})) .
\end{align}
Here $\bar n$ is the mean galaxy number density.
In particular, in the linear bias model one has
\begin{align}
\lambda_{\rm gal}(\bm{x})
=
\bar n
\left[
1+b_1\delta(\bm{x})
\right] ,
\end{align}
and the galaxy two-point correlation function becomes
\begin{align}
\xi_{\rm gal}(r)
=
b_1^2\,\xi_{\rm m}(r)
\end{align}
\citep{1984ApJ...284L...9K}.
More generally, one may expand the local nonlinear bias as
\begin{align}
\lambda_{\rm gal}(\bm{x})
=
\bar n
\left[
1
+
b_1\delta(\bm{x})
+
\frac{b_2}{2}\delta^2(\bm{x})
+
\cdots
\right]
\end{align}
\citep{1993ApJ...413..447F}.
\CG{More general bias prescriptions may also include dependence on smoothed density fields, tidal fields, and environmental variables \citep{2018PhR...733....1D}.}
\CG{The effect of galaxy bias therefore appears through modifications of the correlation structure inherited from the underlying matter distribution.}

\CG{\subsection{Bias and stochastic sampling}}

\CG{Given the galaxy formation intensity $\lambda_{\rm gal}(\bm{x})$, the observed catalog is obtained through stochastic point-process sampling.}
In the simplest case,
\begin{align}
X
\sim
\mathrm{PPP}
\!\left(
\lambda_{\rm gal}(\bm{x})\,\pd^3x
\right)
\end{align}
holds, where PPP denotes a Poisson point process.

More generally, one may also describe the galaxy distribution as a Cox process, in which the intensity field itself is random \citep[e.g.,][]{2003Daley_point_processI}.
\CG{The sampling process introduces an additional source of stochasticity beyond the transport and bias components of structure formation.}
\CG{Consequently, the Wasserstein distance between the initial density field and the observed galaxy catalog contains contributions arising from gravitational transport, galaxy bias, and finite sampling.}
\begin{align}
W_2(\rho_{\rm lin},\mu_N) .
\end{align}
\CG{In the following sections we quantify how galaxy bias and point-process sampling modify the transport distance through correlation and shot-noise corrections.}
\\

\section{Corrections to the Wasserstein distance for correlated point processes}
\label{sec:correlated_process_wasserstein}

In the previous section we introduced the effects of galaxy bias and stochastic sampling.
We now derive the correction to the Wasserstein distance arising from spatial correlations in the point process.

\subsection{Correlation correction}

\begin{figure}
\centering\includegraphics[width=\linewidth]{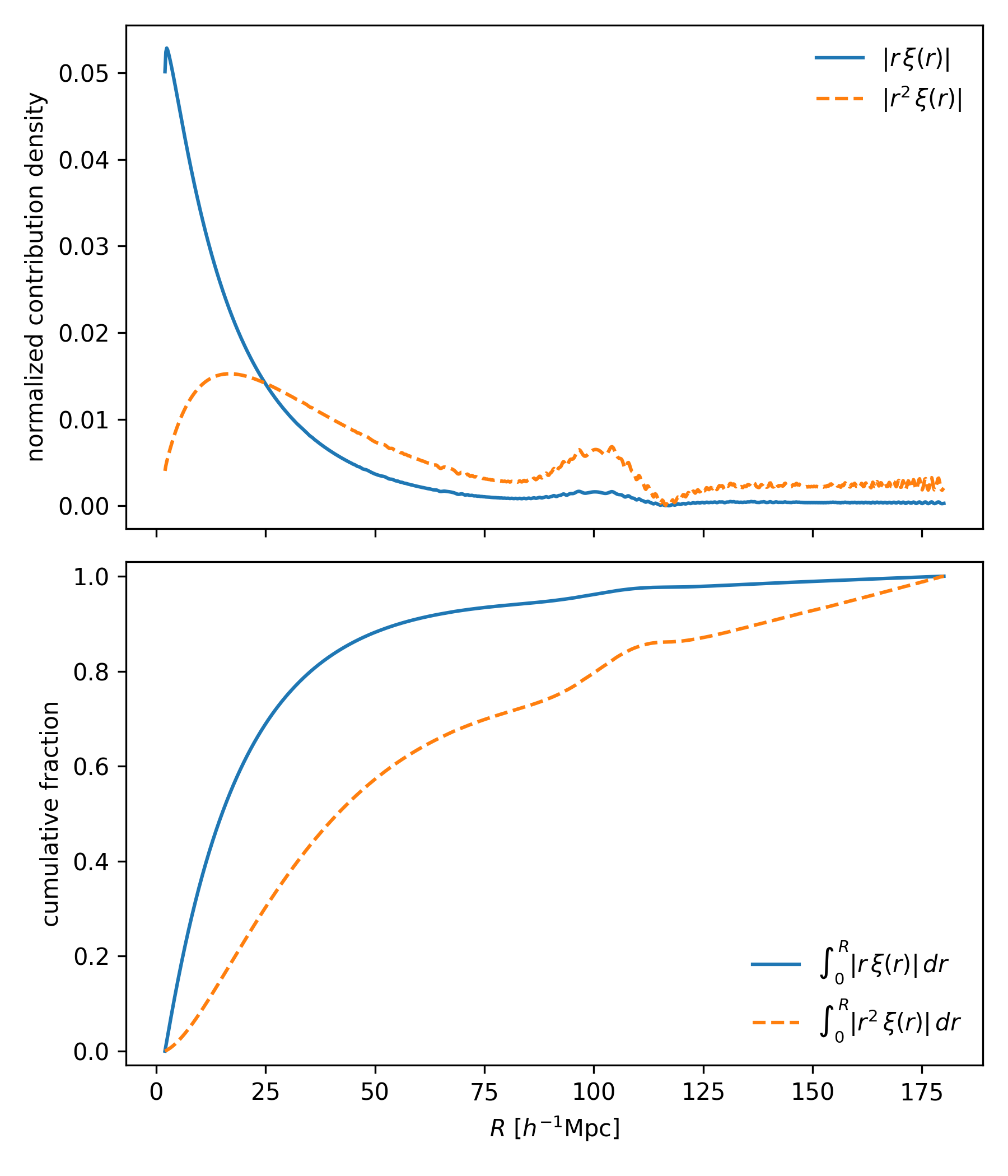}
\caption{\CG{Comparison between the transport-weighted kernel $r\xi(r)$ and the conventional volume-weighted kernel $r^2\xi(r)$ for a linear $\Lambda$CDM correlation function. 
The upper panel shows the normalized contribution density of each kernel as a function of separation. 
The lower panel shows the corresponding cumulative integrals. 
}
}\label{fig:wasserstein_traditional_comparison}
\end{figure}

For a homogeneous point process with mean number density $\bar n$, the two-point density is given by
\begin{align}
\rho^{(2)}(\bm{x},\bm{y})
=
\bar n^2
\left[
1+\xi(\bm{x}-\bm{y})
\right] .
\end{align}
Here $\xi$ denotes the two-point correlation function.
Assuming statistical isotropy, one may write
\begin{align}
\xi(\bm{x}-\bm{y})
=
\xi(|\bm{x}-\bm{y}|) .
\end{align}
To express the correlation structure in Fourier space, we introduce the structure factor
\begin{align}
S(\bm{k})
=
1+\bar n\,\widetilde\xi(\bm{k}) ,
\end{align}
where $\widetilde\xi(\bm{k})$ is the Fourier transform of $\xi(\bm{r})$.
The power spectrum of the point process can then be decomposed as
\begin{align}
P_{\rm pp}(k)
=
\frac{1}{\bar n}
+
\widetilde\xi(k) .
\end{align}
Using the same linearized approximation as in Sec.~\ref{sec:density_fluctuation_wasserstein}, one obtains
\begin{align}
\expt[W_2^2]
\simeq
\frac{1}{2\pi^2}
\int_0^\infty
P_{\rm pp}(k)
\,\pd k .
\end{align}
Therefore the expectation value of the Wasserstein distance can be decomposed as
\begin{align}
\expt[W_2^2]
=
W_{2,\rm shot}^2
+
\Delta W_{2,\rm corr}^2 .
\end{align}
The correlation correction is thus given by
\begin{align}
\Delta W_{2,\rm corr}^2
=
\frac{1}{2\pi^2}
\int_0^\infty
\widetilde\xi(k)
\,\pd k .
\end{align}
Under isotropy, using the Fourier transform relation
\begin{align}
\widetilde\xi(k)
=
4\pi
\int_0^\infty
\xi(r)
\frac{\sin kr}{kr}
r^2
\,\pd r ,
\end{align}
one obtains
\begin{align}
\Delta W_{2,\rm corr}^2
=
\int_0^\infty
r\,\xi(r)
\,\pd r . \label{eq:wasserstein_corr_realspace}
\end{align}
\CG{The appearance of the kernel $r\,\xi(r)$ constitutes the central result of this work.}

\CG{\subsection{Physical interpretation}}
\CG{The scale dependence of the transport-weighted kernel $r\xi(r)$ and the conventional volume-weighted kernel $r^2\xi(r)$ is illustrated in Fig.~\ref{fig:wasserstein_traditional_comparison}.
The curves are computed from the linear-theory matter correlation function of a fiducial $\Lambda$CDM cosmology \citep{1998ApJ...496..605E} , using the Planck 2018 parameter set \citep{2020A&A...641A...6P}.
The figure shows that the two weightings probe substantially different physical scales.
In Fig.~\ref{fig:wasserstein_traditional_comparison}, the transport-weighted kernel (upper panel) is dominated by small and intermediate scales, whereas the volume-weighted kernel receives substantial contributions from larger separations and exhibits a pronounced enhancement around the baryon acoustic oscillation (BAO) scale. 
In contrast, the transport-weighted integral (lower panel) converges rapidly within $\sim 50\text{--}60\,h^{-1}\mpc$, while the volume-weighted integral continues to accumulate contributions up to the BAO scale and beyond. 
This demonstrates that Wasserstein-based statistics probe a scale dependence distinct from that of conventional clustering measures.}

\CG{The weighting appearing in Eq.~(\ref{eq:wasserstein_corr_realspace}) differs fundamentally from that of conventional clustering statistics.}
\CG{In standard correlation analyses, the factor $r^2$ arises from the volume element of spherical shells, so that quantities of the form}
\begin{align}
\int_0^\infty
r^2\xi(r)
\,\pd r
\end{align}
\CG{measure the contribution of galaxy pairs at different separations.}

\CG{By contrast, the Wasserstein correction is weighted linearly in separation,}
\begin{align}
\Delta W_{2,\rm corr}^2
=
\int_0^\infty
r\,\xi(r)
\,\pd r .
\end{align}
\CG{This weighting reflects the transport cost associated with moving mass elements over a distance $r$.}
\CG{The Wasserstein distance therefore probes cumulative mass displacements rather than pair counts.}

\CG{The difference between the kernels $r\,\xi(r)$ and $r^2\xi(r)$ implies a different sensitivity to spatial scales.}
\CG{In particular, the transport weighting enhances the role of coherent displacements while suppressing the purely volumetric weighting characteristic of conventional clustering statistics.}

\begin{figure}
\centering
\includegraphics[width=\linewidth]{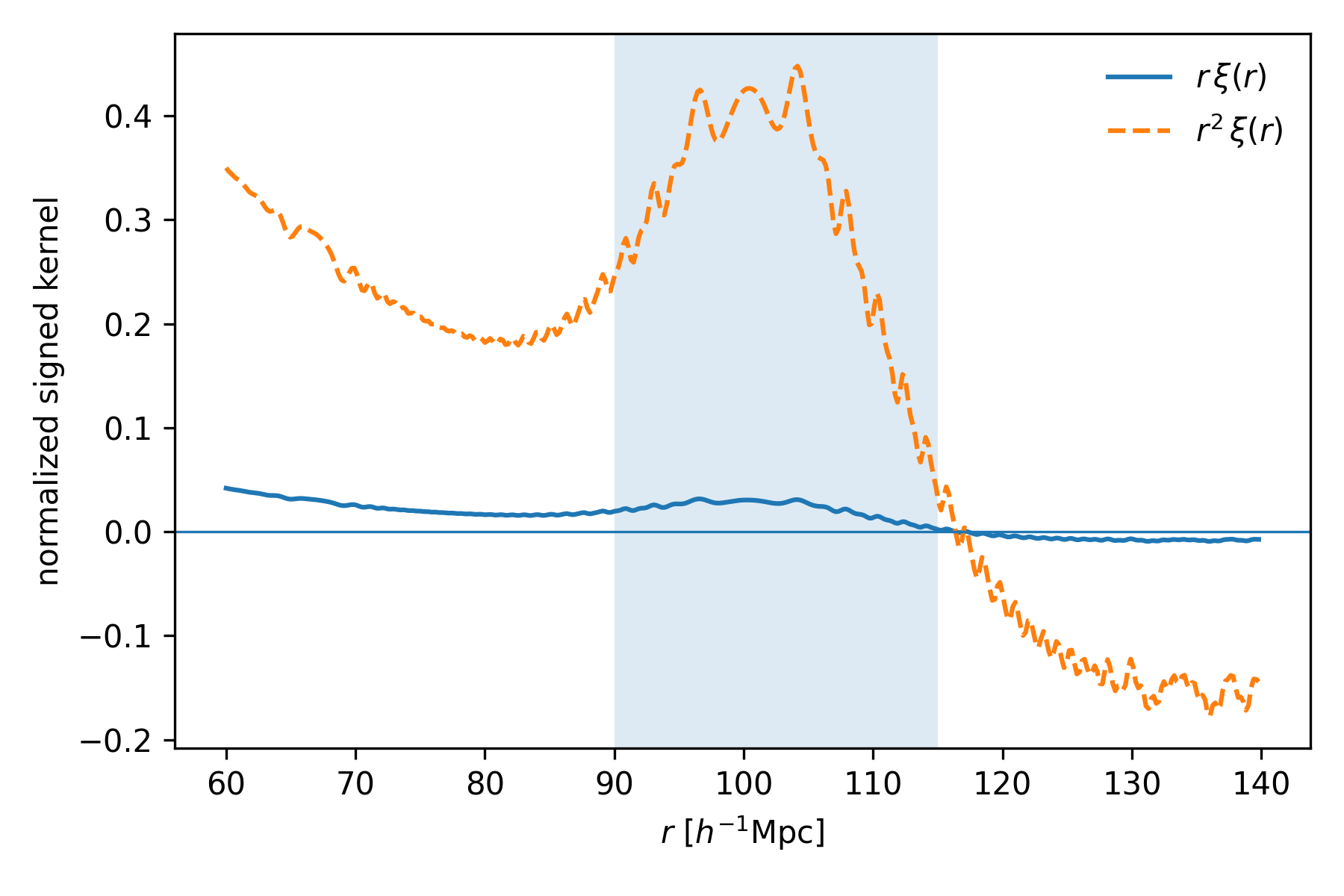}
\caption{
\CG{Comparison of the transport-weighted kernel $r\xi(r)$ and the conventional volume-weighted kernel $r^{2}\xi(r)$ around the baryon acoustic oscillation (BAO) scale for a linear $\Lambda$CDM correlation function. The shaded region indicates the approximate BAO range.} 
}\label{fig:bao_weighting}
\end{figure}

\CG{This distinction becomes especially relevant around the baryon acoustic oscillation (BAO) scale, where large-scale coherent motions broaden the acoustic feature through the displacement of galaxies.}
\CG{Because the Wasserstein distance is directly connected to the displacement field, it naturally probes the transport processes responsible for BAO smearing.
Figure~\ref{fig:bao_weighting} illustrates this difference in more detail around the BAO scale. Although the acoustic feature appears prominently in the conventional volume-weighted kernel $r^{2}\xi(r)$, its contribution to the transport-weighted kernel $r\xi(r)$ is substantially weaker. 
Combined with the cumulative integrals shown in Fig.~\ref{fig:wasserstein_traditional_comparison}, this indicates that the transport-weighted quantity is dominated primarily by small- and intermediate-scale displacements, whereas the volume-weighted statistic continues to accumulate significant contributions up to the BAO scale and beyond. 
The Wasserstein correction therefore probes the cumulative rearrangement of matter rather than the volume-weighted abundance of galaxy pairs.
}

\CG{\subsection{Finite-volume correction}}

In a finite volume the minimum wave number
\begin{align}
k_{\min}
=
\frac{2\pi}{L}
\end{align}
exists, and therefore
\begin{align}
\Delta W_{2,\rm corr,V}^2
=
\frac{1}{2\pi^2}
\int_{k_{\min}}^{k_{\max}}
\widetilde\xi(k)
\,\pd k .
\end{align}
In real space this can be written as
\begin{align}
\Delta W_{2,\rm corr,V}^2
=
\int_0^\infty
r\,\xi(r)\,
\mathcal K_L(r)
\,\pd r , \label{eq:suppression_function_obs}
\end{align}
where $\mathcal K_L(r)$ is a suppression function determined by the finite observational window.

For galaxy distributions with positive clustering one has
\begin{align}
\Delta W_{2,\rm corr}^2
>
0 ,
\end{align}
so that the transport distance increases relative to the Poisson reference model.
\CG{Consequently, finite-volume effects suppress the contribution of long-wavelength correlations to the transport distance in the same way that they suppress large-scale displacement modes.}

\CG{The quantity}
\begin{align}
\Delta W_{2,\rm corr}^2
\sim
\int_0^\infty
r\,\xi_{\rm gal}(r)
\,\pd r
\end{align}
\CG{therefore provides a transport-based measure of galaxy clustering that is distinct from conventional pair-count statistics.}

\section{Galaxy formation bias and a unified Wasserstein theory}
\label{sec:unified_galaxy_bias_wasserstein}

In the previous section we showed that the correction to the Wasserstein distance for a correlated point process is governed by
\begin{align}
\Delta W_{2,\rm corr}^2
\sim
\int_0^\infty
r\,\xi(r)\,\pd r .
\end{align}
\CG{Using the bias expansion introduced in Sec.~\ref{sec:galaxy_formation_as_point_process}, this result can be translated into an explicit correction associated with galaxy formation.}

\subsection{Galaxy formation bias}

\CG{For the linear bias model}
\begin{align}
\delta_{\rm g}(\bm{x})
=
b_1\,\delta(\bm{x}),
\end{align}
the galaxy correlation function becomes
\begin{align}
\xi_{\rm gal}(r)
=
b_1^2\,\xi_{\rm m}(r) .
\end{align}
Therefore
\begin{align}
\Delta W_{2,\rm gal}^2
=
b_1^2
\int_0^\infty
r\,\xi_{\rm m}(r)\,\pd r ,
\end{align}
so that the Wasserstein correction is simply rescaled by the factor $b_1^2$.
\CG{For nonlinear bias models,}
\begin{align}
\delta_{\rm g}(\bm{x})
=
b_1\delta(\bm{x})
+
\frac{b_2}{2}\delta^2(\bm{x})
+
\cdots ,
\end{align}
the correlation function becomes
\begin{align}
\xi_{\rm gal}(r)
=
b_1^2\xi_{\rm m}(r)
+
b_1b_2
\left<
\delta(\bm{x})
\delta^2(\bm{x}+\bm{r})
\right>
+
\cdots .
\end{align}

\CG{For Gaussian initial conditions, this term should be understood within standard perturbation theory rather than as a purely Gaussian three-point function.}
\CG{Writing $\delta=\delta^{(1)}+\delta^{(2)}+\cdots$, the tree-level Gaussian contribution $\langle\delta^{(1)}(\bm{x})[\delta^{(1)}(\bm{x}+\bm{r})]^2\rangle$ vanishes.}
\CG{The leading non-vanishing contribution instead arises from nonlinear mode coupling, such as $\langle\delta^{(2)}(\bm{x})[\delta^{(1)}(\bm{x}+\bm{r})]^2\rangle$, which scales as $\sigma^2\xi_{\rm m}(r)$.
We therefore write, }
\begin{align}
\left<
\delta(\bm{x})
\delta^2(\bm{x}+\bm{r})
\right>
\simeq
2\sigma^2
\xi_{\rm m}(r) .
\end{align}
Hence
\begin{align}
\xi_{\rm gal}(r)
=
\left(
b_1^2
+
2b_1b_2\sigma^2
+
\cdots
\right)
\xi_{\rm m}(r),
\end{align}
and therefore
\begin{align}
\Delta W_{2,\rm gal}^2
=
\left(
b_1^2
+
2b_1b_2\sigma^2
+
\cdots
\right)
\int_0^\infty
r\,\xi_{\rm m}(r)\,\pd r .
\end{align}

More general bias prescriptions involving tidal and environmental fields lead to
\begin{align}
\xi_{\rm gal}(r)
=
b_1^2\xi_{\rm m}(r)
+
2b_1b_s\xi_{\delta s}(r)
+
b_s^2\xi_s(r)
+
\cdots ,
\end{align}
and consequently
\begin{align}
\Delta W_{2,\rm gal}^2
=
\int_0^\infty
r
\left[
b_1^2\xi_{\rm m}(r)
+
2b_1b_s\xi_{\delta s}(r)
+
b_s^2\xi_s(r)
+
\cdots
\right]
\pd r .
\end{align}
\CG{Regardless of the specific bias model, the Wasserstein correction is governed by the geometrically weighted kernel $r\,\xi(r)$.}

\subsection{Unified theoretical expression}

\CG{Combining the results of Secs.~\ref{sec:density_fluctuation_wasserstein}, \ref{sec:correlated_process_wasserstein}, and \ref{sec:galaxy_formation_as_point_process}, the Wasserstein distance may be decomposed into transport, bias, and sampling contributions,}
\begin{align}
W_2^2
\simeq
W_{\rm grav}^2
+
W_{\rm bias}^2
+
W_{\rm samp}^2 .
\end{align}
\CG{The bias contribution is}
\begin{align}
W_{\rm bias}^2
\sim
\int_0^\infty
r
\left[
\xi_{\rm gal}(r)
-
\xi_{\rm m}(r)
\right]
\pd r .
\end{align}
\CG{The shot-noise contribution arising from point-process sampling is}
\begin{align}
\left<
W_{\rm samp}^2
\right>
\simeq
\frac{1}{2\pi^2\bar n}
\int_0^\infty
|\widetilde W_R(k)|^2
\,\pd k ,
\end{align}
which becomes
\begin{align}
\left<
W_{\rm samp}^2
\right>
=
\frac{1}{4\pi^{3/2}\bar n R}
\end{align}
for Gaussian smoothing.

\CG{Substituting the transport, bias, and shot-noise terms into the above decomposition yields}
\begin{widetext}
\begin{align}
\left<
W_2^2(\rho_{\rm lin},\mu_N)
\right>
\simeq
\frac{1}{2\pi^2}
\int
P_{\rm m}(k)\,\pd k
+
\int_0^\infty
r
\left[
\xi_{\rm gal}(r)
-
\xi_{\rm m}(r)
\right]
\pd r
+
\frac{1}{2\pi^2\bar n}
\int_0^\infty
|\widetilde W_R(k)|^2
\,\pd k .
\end{align}
\end{widetext}

\CG{Using the finite-volume corrections (\ref{eq:suppression_function_obs}) derived in Sec.~\ref{sec:finite_volume_effect}, one obtains}
\begin{widetext}
\begin{align}
\left<
W_{2,V}^2(\rho_{\rm lin},\mu_N)
\right>
\simeq
\frac{1}{2\pi^2}
\int_{k_{\min}}^{k_{\max}}
P_{\rm m}(k)\,\pd k
+
\int_0^\infty
r\,\xi_{\rm gal}(r)
\mathcal K_L(r)
\,\pd r
+
\frac{1}{2\pi^2\bar n}
\int_{k_{\min}}^{k_{\max}}
|\widetilde W_R(k)|^2
\,\pd k . \label{eq:final_formula_finite_volue}
\end{align}
\end{widetext}
\CG{This expression constitutes the main theoretical result of the paper. The first term represents gravitational transport, the second describes the correlation contribution of the observed galaxy distribution, and the third corresponds to shot noise generated by finite sampling.}
\CG{We note that the second term in Eq.~\eqref{eq:final_formula_finite_volue} corresponds to the finite-volume form of Eq.~\ref{eq:suppression_function_obs} applied to the observed galaxy point process.
The corresponding bias correction is obtained by replacing $\xi_{\rm gal}(r)$ with $\xi_{\rm gal}(r)-\xi_m(r)$.}

\CG{The Wasserstein distance therefore provides a unified transport-based statistic that simultaneously captures mass rearrangement, galaxy formation, and observational discreteness within a single geometric framework.}

\section{Discussion and Conclusions}
\label{sec:discusssion_conclusion}

\CG{
In this work, we formulated cosmological structure formation as a transport problem and investigated the Wasserstein distance between the initial density field and the observed galaxy catalog. 
Under the small-fluctuation approximation, we showed that the Wasserstein distance can be decomposed into contributions associated with gravitational transport, galaxy formation, and stochastic sampling. 
This leads to a unified expression connecting the matter power spectrum, galaxy correlation function, and shot-noise corrections within a common transport-geometric framework.
}

\CG{
A central result is that the correlation contribution to the Wasserstein distance is governed by the transport-weighted kernel
\begin{align}
\Delta W_2^2 \sim \int_0^\infty r \xi(r)\,\pd r .
\end{align}
Unlike conventional volume-weighted clustering measures, this quantity is directly related to cumulative mass displacements and therefore probes a distinct aspect of structure formation. 
The comparison with conventional correlation statistics presented in Sec.~\ref{sec:correlated_process_wasserstein} demonstrates that the transport weighting emphasizes different physical scales and naturally connects the Wasserstein distance to displacement-driven phenomena such as BAO smearing.
}

\CG{
Several important extensions remain for future work. These include the treatment of fully nonlinear transport beyond the Zel'dovich approximation, the incorporation of shell crossing and multistreaming, a more complete treatment of correlated sampling processes, and the construction of practical estimators applicable to realistic galaxy surveys. 
It will also be important to investigate quantitative connections between Wasserstein statistics and observable quantities such as BAO damping scales, reconstruction performance, and other transport-related signatures of large-scale structure formation.
}

\CG{
More broadly, the present work suggests that cosmological structure formation may be viewed as a geometric evolution in the space of probability measures. Within this perspective, gravitational evolution, galaxy formation, and observational sampling appear as successive modifications of a transport distance. Extending this framework beyond the linear regime may provide a route toward a more general transport-geometric description of cosmic structure formation.
}

\section*{Acknowledgments}

\CG{We deeply thank the referee for the careful reading of the manuscript and for many valuable comments and suggestions.}
We are deeply grateful to Shiro Ikeda for valuable discussions and helpful insights that initiated this work.
This work was supported by JSPS Grant-in-Aid for Scientific Research (24H00247) and by the Joint Research Program of the Institute of Statistical Mathematics (General Research 2), ``Machine-learning cosmogony: from structure formation to galaxy evolution.''

\bibliographystyle{apsrev4-2}
\bibliography{optimal_transport}% Produces the bibliography via BibTeX.

\end{document}